\documentclass[graybox]{svmult}
\usepackage{cite}
\usepackage{url}
\usepackage{algorithmic}
\usepackage{algorithm}
\usepackage{graphicx}
\usepackage{xcolor}
\usepackage{verbatim}

\usepackage{type1cm}        % activate if the above 3 fonts are
\usepackage{makeidx}         % allows index generation
\usepackage{graphicx}        % standard LaTeX graphics tool
\usepackage{multicol}        % used for the two-column index
\usepackage[bottom]{footmisc}% places footnotes at page bottom

\usepackage{newtxtext}       % 
\usepackage{newtxmath}       % selects Times Roman as basic font

\makeindex   
\begin{document}

\title*{Resting-state EEG sex classification using selected brain connectivity representation\\
}
\author{Jean Li, Jeremiah D. Deng, Divya Adhia and Dirk de Ridder}
\institute{Jean Li \at Department of Inforamtion Science, University of Otago, \email{jean.li@postgrad.otago.ac.nz}
\and 
Jeremiah D. Deng \at Department of Inforamtion Science, University of Otago, \email{jeremaih.deng@otago.ac.nz}
\and 
Divya Adhia \at Department of Surgical Science, University of Otago, \email{divya.adhia@otago.ac.nz}
\and
Dirk de Ridder \at Department of Surgical Science, University of Otago, \email{dirk.deridder@otago.ac.nz}}

\maketitle

\abstract*{
Effective analysis of EEG signals for potential clinical applications remains a challenging task. So far, the analysis and conditioning of EEG have largely remained sex-neutral. 
This paper employs a machine learning approach to explore the evidence of sex effects on EEG signals, and confirms the generality of these effects by achieving successful sex prediction of resting-state EEG signals. We have found that the brain connectivity represented by the coherence between certain sensor channels are good predictors of sex.
}

\section{Introduction}
Electroencephalography (EEG) is a widely used non-invasive technique to measure multi-channel potentials that reflect the electrical activity of the brain. Over the last a few decades, EEG analysis has been an intensively explored research topic due to its potentials in being applied to the diagnosis of neurological diseases, such as epilepsy, brain tumors, head injury, sleep disorders, dementia, etc.~\cite{PJL:10}. Despite many advances made in recent years, EEG signal analysis remains a challenging task. In addition to being non-stationary, EEG signals often have high noise-to-information ratios, and they can be significantly affected by various artifacts, demonstrating characteristics that differ from signals generated by activities in the brain~\cite{Vanneste:18}. Common artifacts include eye movements, jaw tension, and muscle contractions. To make effective signal analysis even more challenging, EEG signals are highly individual-specific, and cross-subject pattern identification can be elusive. 

In a more proactive approach, EEG can also be applied to biofeedback training as an operant conditioning technique to reinforce or inhibit specific forms of EEG activities. It has been used in anxiety and addiction treatment, also employed for attentional, cognitive, and psychosocial functioning improvement~\cite{Scott:05}. It is noted that popular EEG biofeedback treatment is largely based on sex-neutral protocols~\cite{William:2005}. Our proposition is that if there are innate sex differences found in EEG signals, then it is possible for new sex-differentiated EEG biofeedback protocols to be developed to potentially enhance many neurological treatments.

This study focuses on using machine learning techniques to explore evidence of sex effects on EEG signals. Rather than just examining the between-sex statistics differences, we attempt to construct an effective classification model to predict the sex of a subject through their EEG signals. 

This chapter is organized as follows. In Section \ref{sec:RW}, we review related work and discuss some common approaches to represent EEG signals. Section \ref{sec:data} introduces the data set used in this study, and the data pre-processing procedures, feature extraction and selection, and the classification methods. Then in Section \ref{sec:results}, we present the results of sex classification. We conclude the paper in Section \ref{CCL}.

\section{Related work}\label{sec:RW}
Within the neuroscience literature, there has been an ongoing interest in sex effects in cognitive performance and the underlying neural mechanisms~\cite{Hirnstein:18}. While a number of studies failed to find sex differences in cognitive performance and hemispheric asymmetry~\cite{Beste:10,Papousek:11}, 
some evidence of the sex effects on EEG has been found. 

In~\cite{Clarke:18}, the EEG signals of 80 individuals between the ages of 8 and 12 years were analyzed. Differences between sex were found in this study group, with males having less theta but more alpha frequency components than females. Females were also found to have a developmental lag in the EEG compared to males. An earlier work~\cite{Davidson:76} also reported sex differences in EEG asymmetry during self-generated cognitive and affective tasks. However, both of these studies did not validate the findings on external test individuals, therefore, the generality of the findings may be limited.

The study in~\cite{Carrier:01} investigated the effects of age and sex on sleep EEG power spectral density of individuals of age ranged from 20–60 years. The average power density within the 4-second epochs was calculated. It was found that females show significantly higher spectral power density in some power bands than males. Though significant effects of age on sleep EEG spectral power density were found, the study did not find any interaction between age and sex. This study performed robust statistical analysis for longitudinal data. However, no external data validation is performed to demonstrate the generalization ability on new subjects. 

Although EEG signal classification has been widely explored for different purposes, there have been few studies that investigate sex differences in EEG and attempt EEG-based sex classification~\cite{Li:20}. For classification purposes in general, the EEG signals are typically pre-processed by band-pass filters and spatial filters for feature extraction. Most commonly, frequency band power features and time point features are employed to represent EEG signals. Band power features represent the average energy level of EEG signals within a certain frequency range over a given time window called an ``epoch''. Band power features need to be extracted respectively in each channel. Time-point features are a concatenation of EEG signals from all channels, and they are typically used for event-related potentials classifications~\cite{Lotte:07}. To cope with the non-stationarity of EEG signals, band power features are usually extracted from a reasonably short epoch. For example, in~\cite{Carrier:01}, epochs are extracted using a sliding window with a length of 2 seconds.

Spatial filters were also applied in other studies for EEG feature extraction. These can be obtained in a supervised manner, such as Common Spatial Patterns (CSP). CSP projects the signals into another matrix space that maximizes the distance between 2 classes.  Reference~\cite{Lotte:07} discusses the effectiveness of this approach and has proven it to be useful. Spatial filters can also be obtained through an unsupervised way such as Independent Component Analysis (ICA). In addition to the above, other EEG representing methods are also studied, these include sparse representation and deep learning. The sparse representation-based classification (SRC) method has shown a robust classification performance~\cite{Shin:15}. Deep learning, in which the features and the classifier are jointly learned directly from the EEG signals. The convolutional neural networks and restricted Boltzmann machines are the two most popular deep learning methods for EEG-based Brain-Computer Interfaces (BCIs) studies~\cite{Lotte:07}. 

Correlation between EEG channels, also known as coherence, was evaluated in time-domain to analyze the connectivity patterns in dystonia patients~\cite{Baltazar:20}. Coherences have the advantage of yielding the possibility to recognize motor-imagery related activation even without typical activation observed, often giving small standard deviations~\cite{Markovic:20}. To better handle non-stationarity, connectivity has been modeled by coherence obtained from spectral features obtained from FFT~\cite{MURIAS:07,Hoeller:13,Mumtaz:17,Markovic:20}. Connectivity based on spectral coherence is also found to be an effective biometric feature~\cite{LaRocca:14}.  

More recently,~\cite{Putten:2018} utilized deep learning to predict sex through EEG signals. With a large data size (1000 adults) and deep convolutional nets, an accuracy of 81\% was achieved. It shows that the beta band provides the most important features in predicting sex. Another deep-learning based study~\cite{Zhang:19} assesses gender differences in emotion processing EEG data and reports a classification accuracy as high as 95\% using gamma band features. The dataset contains however only 60 subjects. 

\section{Data and methods}\label{sec:data}
%In this section, we will introduce the EEG dataset used for this study, and present the methods used in this study.

\subsection{Data set description}
The dataset used for this study is part of the data collected for a previous work~\cite{Vanneste:18}, containing a raw resting-state EEG streams set of 241 healthy individuals only. The raw EEG signal was collected through a standard Mitsar amplifier with 19 channels (Fp1, Fp2, F7, F3, Fz, F4, F8, T7, C3, Cz, C4, T8, P7, P3, Pz, P4, P8, O1, O2) with a sampling rate of 250 Hz. The studied population consists of 150 females and 91 males aged between 17 to 89 years. The distribution regarding the age and sex of the population can be found in Table~\ref{tablepopulation}. Note that there is a notable imbalance between two sexes across several age groups.

\begin{table}[htbp]
\caption{Age and sex distribution of the sample set}
\begin{center}
\begin{tabular}{|p{2cm}|p{2cm}|p{2cm}|}
\hline
\textbf{Age (years)} & \textbf{\textit{Females}}& \textbf{\textit{Males}} \\
\hline
17-25& 35 & 20  \\
\hline
26-35& 17 & 9  \\
\hline
36-45 & 36 & 19 \\
\hline
46-55 & 34 & 21 \\
\hline
56-65 & 16 & 15 \\
\hline
66-75 & 8 & 5 \\
\hline
76-89 & 4 & 2 \\
\hline
Sum & 150 & 91 \\
\hline
\end{tabular}
\label{tablepopulation}
\end{center}
\end{table}

\subsection{Preprocessing}
Since EEG signals cannot be segmented into physiologically relevant units, the conventional approach of segmenting the EEG streams into epochs according to time interval is adopted in this study. In this study, we segmented the raw EEG streams into 2-second epochs. The first 5 seconds of every EEG recording were discarded to avoid possible noise.

\subsection{Signal representation}

In this study, the raw EEG signals of each subject are represented by the spectral connectivity between channel pairs. 
The MNE package\footnote{Available from \url{https://mne.tools/stable/generated/mne.connectivity.spectral_connectivity.html}} was applied to compute the frequency-domain connectivity measures.
In particular, for every pair of channels, the coherence across all epochs in each frequency was computed, as in Eq.~\ref{eq:coh}.

\begin{equation}\label{eq:coh}
    C=\frac{| E[S_{xy}] |}{\sqrt{ E[S_{xx}] \times E[S_{yy}] }}
\end{equation}
where $S_{xy}$ is the cross-spectral density between $x$ and $y$, and $E[]$ denotes average over epochs.\\

To shorten the feature vector length, the mean value of the connectivity within 5 major brain wavebands are adopted.
These frequency bands are: 0–4 Hz (delta), 4–8 Hz (theta), 8–13 Hz (alpha), 13–30 Hz (beta) and 30-45 Hz (gamma). 
This gives us a feature vector of length 855 (171 channels pairs $\times$ 5 bands).

\subsection{Feature analysis}\label{sec:dimension}

Compared with the relatively small sample size of 241, the feature vector of length 855 may cause a potential over-fitting problem. To further shorten the connectivity representation, training subjects splitting and feature selection were adopted.

For each training subject, we split the subject's entire EEG recordings into several 30 seconds long (15 epochs) sections, and use each section as an independent training sample. Considering the imbalance of the dataset, we take 5 sections from each female, and 8 from each male to form a more balanced training set. This approach results in a training set with a 5 to 8 times larger sample size.

Feature selection is carried out using XGBoost~\cite{Gramfort:2013}. In each of the 50 trials, We randomly selected 90\% of the subjects, split the recordings as described above, and fit these data points into a XGBoost binary classifier. We then rank all 855 features according to the feature importances given by XGBoost and stack the ranks from the 50 trials together. Since the feature importance ranking varies across trials, we generate our own feature importances through the 855 $\times$ 50 matrix. We iterate from the most to least important XGBoost ranks, the earlier a feature appears in all 50 trials, the more important it is marked. We then select the top 34 connectivity features according to our ranking as the final representation. Detail of the feature selection outcome will be demonstrated in Section~\ref{fs}.

\subsection{Classification}
Four different classifiers are applied separately using the chosen representation. They are XGBoost classifier (XGB), multi-layer perceptron (MLP), support vector machine (SVM), and the random forest classifier (RF).

Validation scores were generated by running 50 trials. In each of the 50 trials, we randomly select 90\% of the subjects, split these recordings as described in Section~\ref{sec:dimension} to form a reasonably balanced training set. The other 10\% of subjects are used for validation. For the subjects in the validation set, EEG streams are not split (i.e. no repeated validation subjects). To increase the stability, we take a longer recording length of 3 minutes (90 epochs) to compute the connectivity and chose the 34 features we selected as the final representations of the validation subjects.

For binary classification, depending on the classification algorithm, a crisp classification decision is made (such as in decision trees), or a 0.5 threshold value is used e.g. in the output node of an MLP. We use a unified probabilistic framework here, by estimating the probability of a classification outcome, and using a relaxed threshold value to decide the classification outcome $c$: 
\begin{equation}
    c=\left\{
    \begin{array}{ll}
        1, & \mathrm{if~} y>\theta \\
        0, &\mathrm{otherwise}
    \end{array}
    \right.,
\end{equation}
where $y$ is the predicted probability, and $\theta$ is the probability threshold which can be optimized through cross-validation. \\

\section{Results}\label{sec:results}

\subsection{Feature selection}\label{fs}

\begin{figure}[t]
\sidecaption
\centerline{\includegraphics[width=0.75\textwidth]{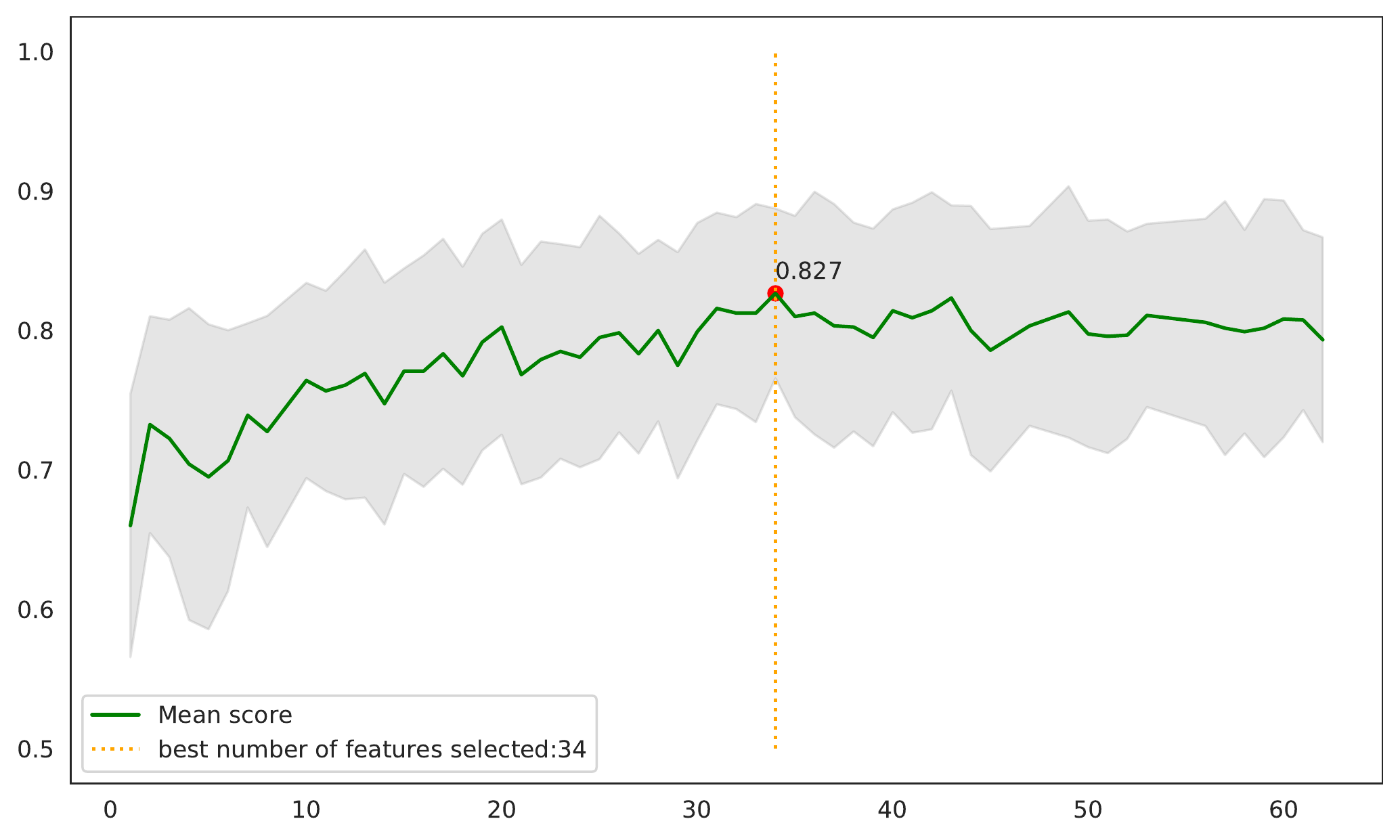}}
\caption{Validation result of XGBoost using increasing numbers of chosen features.}
\label{numf}
\end{figure}

\begin{figure}[t]
\sidecaption
\centerline{\includegraphics[width=1.0\textwidth]{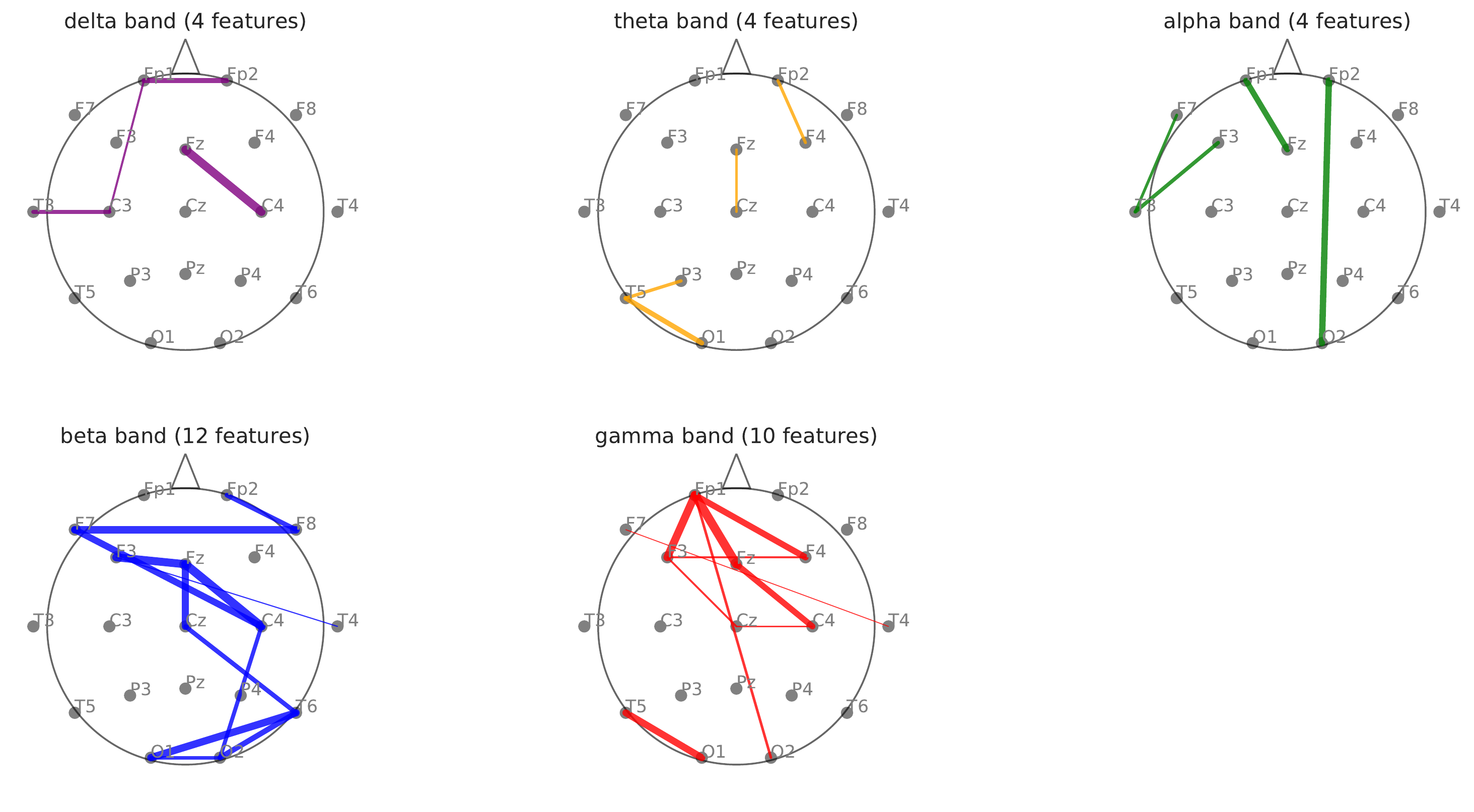}}
\caption{Chosen connectivity features in five frequency bands.
Thicker lines between the channels suggest that the connectivity values between these channels pairs are more important features.}
\label{fig:topo}
\end{figure}

Fig.~\ref{numf} demonstrates the 50 trials validation result from a binary XGBoost classifier using different numbers of features. The classifier reached the best performance when using the top 34 features. As shown in the figure, the mean score displays an increasing trend until the number of features reaches 34. 

\begin{figure}[t]
\sidecaption
\centerline{\includegraphics[width=0.8\textwidth]{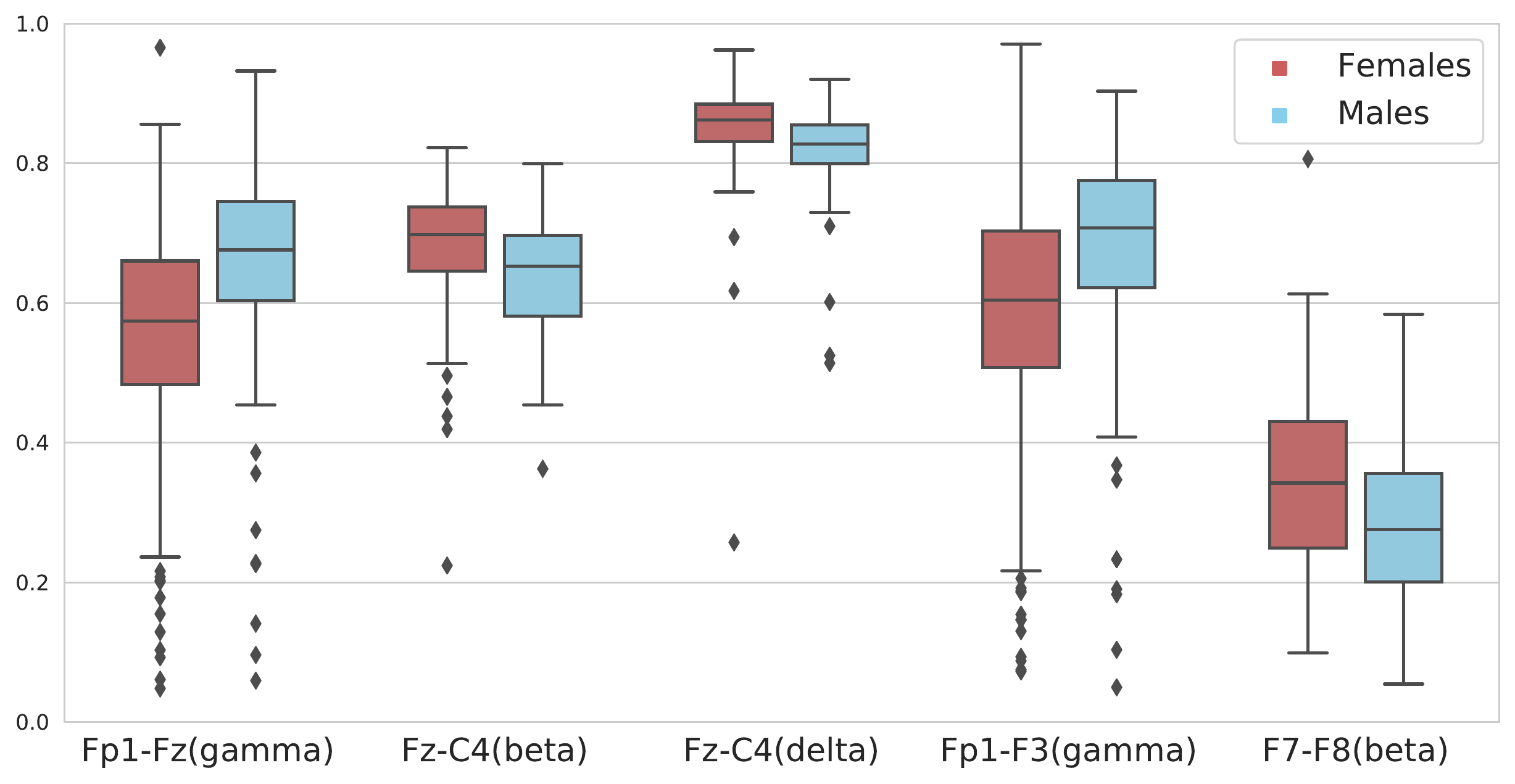}}
\caption{Box plots of some top connectivity features display promising separability of the two sex classes.}
\label{fig:boxplot}
\end{figure}

The chosen 34 features are illustrated in topographical maps in Fig.~\ref{fig:topo}. The brain connectivity in beta and gamma bands seems more prominent than those in other bands when performing sex classification tasks, as more connectivity pairs in these 2 bands are marked as important features. The gamma band connectivity between channel Fp1 and Fz is ranked as the most important feature. The connectivity between channel F7 and F8 in beta band is the 6th important feature. These two important pairs are in agreement with a previous neurosience study~\cite{Ingalhalikar:2014}, in that females have more latero-lateral interhemispheric connectivity (F7-F8) and males more antero-posterior intrahemispheric (Fp1-Fz) connectivity.
The connectivity between channel Fz and C4 appears to be important in three different bands: beta (ranked 2nd), delta (ranked 3rd), and gamma (ranked 13th). We speculate that the Fz-C4 connectivity in beta could be related to the known sex differences in the anterior cingulate cortex (Fz) and sensorimotor area (C4).
Fig.~\ref{fig:boxplot} shows that some of the top connectivity features display promising separability of the two sex classes.

\subsection{Validation results}

\begin{figure}[t]
\sidecaption
\centerline{\includegraphics[width=1\textwidth]{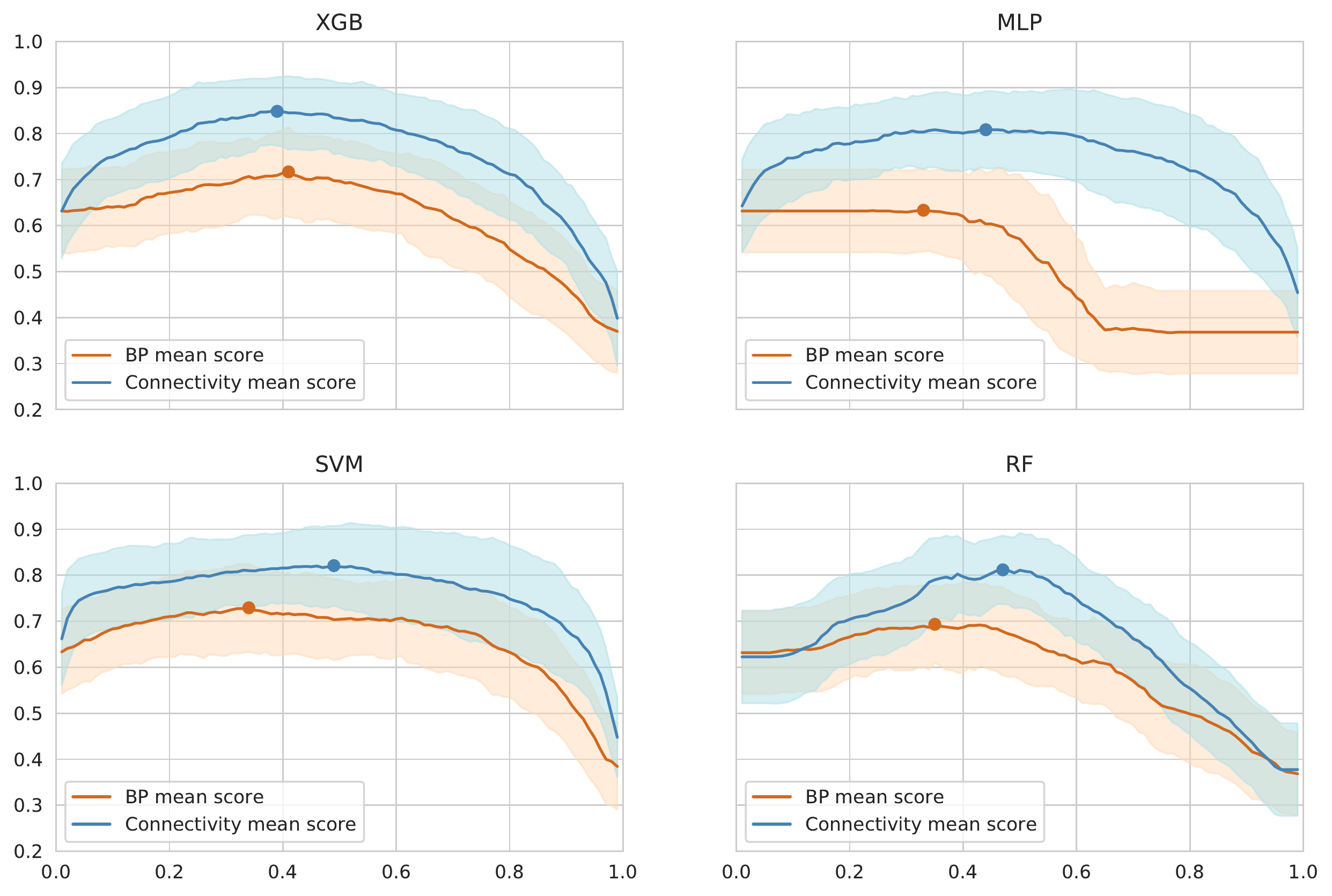}}
\caption{Validation accuracy versus the  threshold ranged in (0, 1). Shaded areas indicate the accuracy values within 1 stdev of the corresponding mean score.}
\label{fig:BPvsConnectivity}
\end{figure}

\begin{figure}[t]
\sidecaption
\centerline{\includegraphics[width=0.8\textwidth]{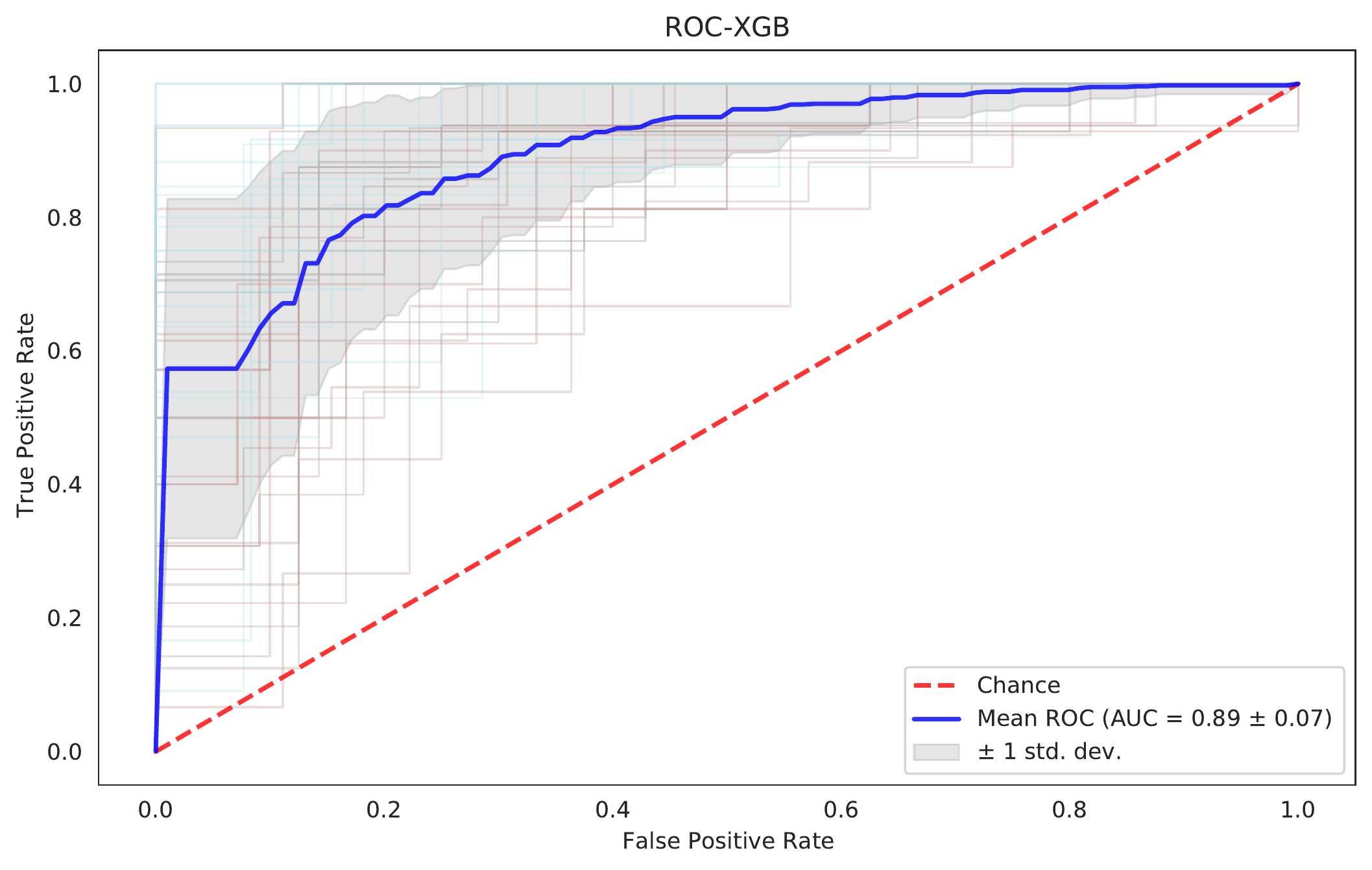}}
\caption{Average ROC curves obtained through cross-validations of XGB.
%~\footnote{original code source: \url{https://scikit-learn.org/stable/auto_examples/model_selection/plot_roc_crossval.html#sphx-glr-auto-examples-model-selection-plot-roc-crossval-py}}
}
\label{fig:AUC_cv}
\end{figure}

Fig.~\ref{fig:BPvsConnectivity} illustrates the 50 trials validation scores across increasing thresholds when using XGB, MLP, SVM, and RF to predict the probability of female class.
The connectivity scores are generated using the chosen 34 connectivity features.
The band power scores are generated by using 31 band power features which are selected in a similar way with the connectivity features. All classifiers demonstrate that using connectivity features outperforms using band power features.
When using the selected connectivity features, all four classifiers reach their best performances when the threshold of the probability is set around 0.4 to 0.5. XGB gives the highest accuracy score of 0.848 using a threshold of 0.39.

The mean AUC of XGB using selected connectivity features across 50 trials is 0.89 ($\pm$0.07), as shown in Fig.~\ref{fig:AUC_cv}. The mean AUCs of other classifiers are SVM: 0.88 ($\pm$0.07), MLP: 0.86 ($\pm$0.08), and RF: 0.88 ($\pm$0.08). The high AUCs from classifiers suggest the generally good quality of the chosen features.

\subsection{Test results}

\begin{figure}[t]
\sidecaption
\centerline{\includegraphics[width=0.9\textwidth]{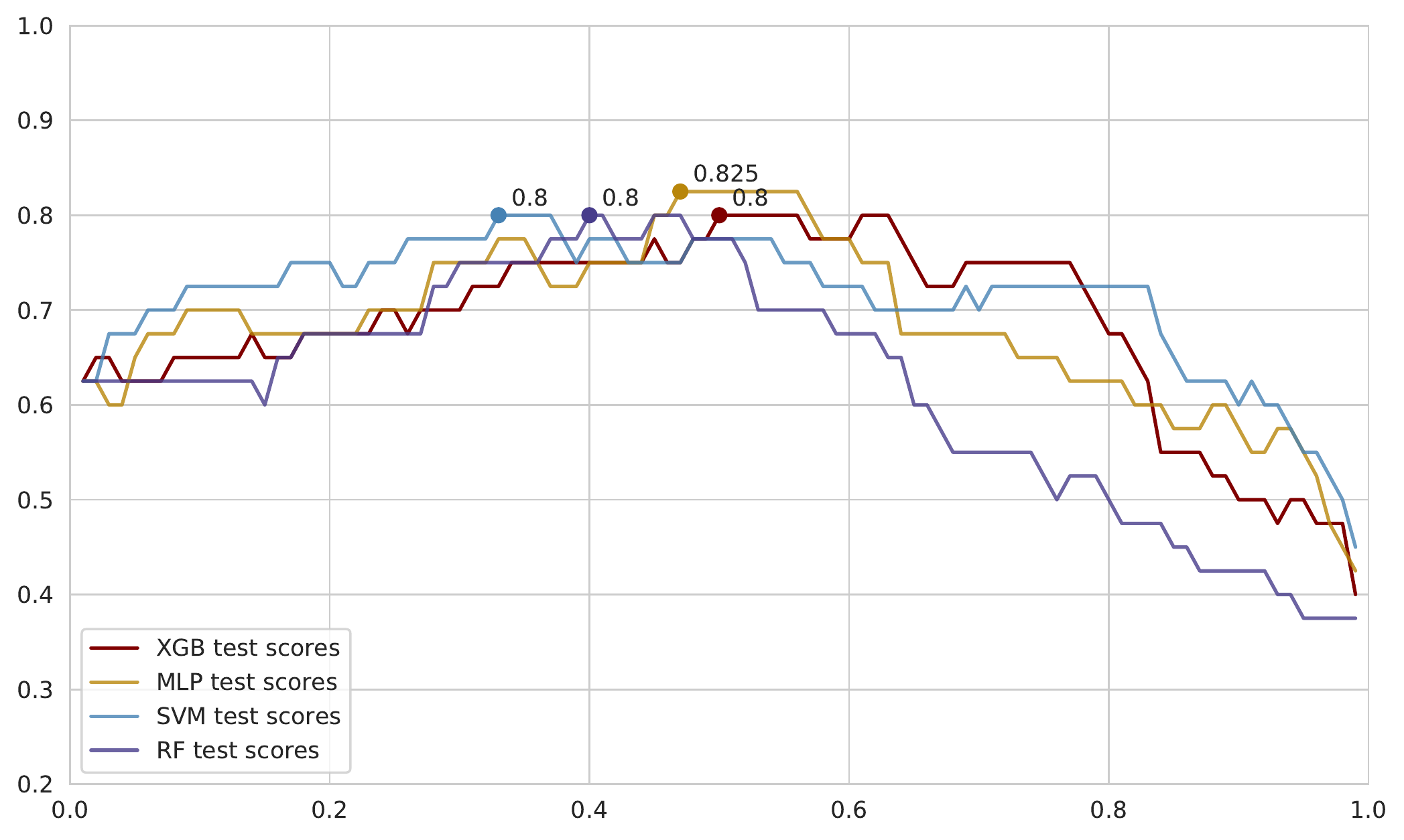}}
\caption{Test accuracy versus the threshold ranged in (0,1). Best accuracy positions indicated by coloured dots.}
\label{thresh_test}
\end{figure}

An independent set comprised of 40 subjects (25 females and 15 males) was used for testing. The test scores across increasing thresholds are shown in Fig.~\ref{thresh_test}. MLP has the best score of 0.825 using a threshold of 0.47, and the best scores of XGB, SVM, and RF are all 0.8.

\section{Conclusions} \label{CCL}
We have conducted a preliminary study on the potential sex difference in resting-state EEG signals using a machine learning approach. Instead of using a data-driven deep learning method which demands more subject data to avoid overfitting, we chose to focus on assessing the effects of band power and connectivity features in sex classification using classifier ensembles and feature analysis methods. In particular, it is found that female and male groups show different brain connectivity patterns which are most prominent in beta and gamma bands. The connectivity between channel Fp1 and Fz in the gamma band shows the greatest sex discrepancy. The connectivity between channel Fz and C4 appears to be different between sexes in the delta, beta, and gamma bands.
The initial band power features are not included in the classification models because they are found to be collectively less important than connectivity features, even though they may contribute positively to the discriminant analysis on sexes. 

Due to limited data availability, we have concentrated on examining healthy subjects' EEG signals. For future work, we would like to investigate sex differences on EEG connectivity with signals obtained under different pathological settings and explore its possible application in biofeedback therapy.

\bibliographystyle{plain}
\bibliography{eeg_sex}

\end{document}